\documentclass[aps,prb,twocolumn,groupedaddress,showpacs]{revtex4}

\usepackage{graphicx}

\begin{document}

\title{Role of electron-electron and electron-phonon interaction effect
in the optical conductivity of VO$_2$}

\author{K.~Okazaki$^1$, S.~Sugai$^1$, Y.~Muraoka$^2$ and Z.~Hiroi$^2$}

\affiliation{$^1$Department of Physics, Nagoya University, Nagoya
464-8602, Japan} 
\affiliation{$^2$Institute for Solid State Physics,
University of Tokyo, Kashiwa 277-8581, Japan}
 
\date{\today}

\begin{abstract}
We have investigated the charge dynamics of VO$_2$ by optical
 reflectivity measurements. Optical conductivity clearly shows a
 metal-insulator transition. In the metallic phase, a broad Drude-like
 structure is observed. On the other hand, in the insulating phase, a
 broad peak structure around 1.3 eV is observed. It is found that this
 broad structure observed in the insulating phase shows a temperature
 dependence. We attribute this to the electron-phonon interaction as in
 the photoemission spectra.
\end{abstract}


\maketitle

\section{Introduction}\label{Intro} 

The metal-insulator transition (MIT) is one of the most interesting
phenomena in the strongly correlated electron systems.~\cite{MIT} VO$_2$
is well known for its first-order metal-insulator transition at 340
K,~\cite{Morin} which is accompanied by a structural transition. In the
high temperature metallic phase it has a rutile structure, while in the
low temperature insulating phase ($M_1$ phase) the V atoms dimerize
along the $c$-axis and the dimers twist, resulting in a monoclinic
structure. The magnetic susceptibility changes from paramagnetic to
nonmagnetic in going from the metallic to the insulating phase. Hence,
this transition is analogous to a Peierls transition. In the early stage
of the study for the MIT in VO$_2$, Goodenough explained the MIT based
on a simplified band model.~\cite{Goodenough} The V $3d$ $t_{2g}$ level
is split into the $d_{\|}$ and $\pi^*$ sublevels due to the tetragonal
crystal field. The $\pi^*$ bands are hybridized with the O 2{\it p}$\pi$
bands and are pushed upward. The $d_{\|}$ band is rather weakly
hybridized with the O $2p$ band and has the lowest energy among the V
3{\it d} bands. In the metallic phase, the $\pi^*$ and $d_{\|}$ bands
overlap and are partially filled. In the insulating phase, the $\pi^*$
bands are shifted upward and the $d_{\|}$ band is split into two
subbands. As a result, the $\pi^*$ bands become empty and the lower
$d_{\|}$ subband is completely filled.~\cite{Goodenough2} However, this
picture was criticized that only with the lattice distortion the optical
gap of 0.6 eV~\cite{Ladd} cannot be reproduced and the importance of
electron correlation effect was pointed out based on the cluster
calculations.~\cite{Sommers} Zylbersztejn and Mott claimed that while
the insulating phase cannot be described correctly without taking into
account of the electron correlation effect, for the metallic phase of
VO$_2$, the screening effect of the $\pi^*$ bands on the $d_{\|}$ bands
is important.~\cite{Zylbersztejn}

On the basis of local-density approximation (LDA) band-structure
calculation, Wentzcovitch {\it et al.}~\cite{Wentzcovitch} concluded
that the insulating phase of VO$_2$ is an ordinary band (Peierls)
insulator. On the other hand, Cr-doped VO$_2$ or pure VO$_2$ under
uniaxial pressure in the [110] direction of the rutile structure has
another monoclinic insulating phase called the $M_2$ phase. In the $M_2$
phase, half of the V atoms form pairs and the other half form zig-zag
chains.~\cite{Marezio} While the V atoms in the pairs are nonmagnetic,
those in the zig-zag chains have local moments and are regarded as
one-dimensional Heisenberg chains according to an NMR
study.~\cite{Pouget} Based on these observations for the $M_2$ phase,
Rice {\it et al.}~\cite{Rice} objected to Wentzcovitch {\it et al} that
the $M_2$ phase is a Mott-Hubbard insulator and the $M_1$ phase also
must be classified as a Mott-Hubbard insulator. More recently, several
theoretical studies have been reported for the MIT of
VO$_2$.~\cite{Tanaka, Biermann, Liebsch, Laad} However, it still remains
highly controversial whether the MIT of VO$_2$ is driven by the
electron-phonon interaction (resulting in a Peierls insulator) or the
electron-electron interaction (resulting in a Mott insulator).

To address this issue, both of the electron-electron and electron-phonon
interaction effects should be further investigated. In a recent
photoemission study, it is concluded that while the electron-electron
interaction is necessary to produce the band gap in the insulating
phase, the electron-phonon interaction is important to fully understand
the electronic structure and charge transport in VO$_2$.~\cite{PES} To
further understand how these effects are related to the charge
transport, it is important to investigate the detailed charge dynamics
of VO$_2$. So far, several optical measurements have been reported for
VO$_2$ using bulk samples~\cite{Ladd, Barker, Verleur} and thin film
samples.~\cite{Verleur, Choi} However, spectroscopic measurements are
difficult for the metallic phase of VO$_2$ with the bulk crystals
because the crystals break into pieces when it cross the
MIT.~\cite{JPSJ} On the other hand, with the thin films, it is difficult
to deduce the optical constants precisely from optical measurements
because multiple reflections between the film and the substrate should
be taken into account.

In this work, we report the optical conductivity of VO$_2$ at various
temperatures deduced from the optical reflectivity measurements of
TiO$_2$ substrates and VO$_2$ thin films grown on the TiO$_2$
substrate. It is observed that a broad Drude-like component in the
metallic phase transfers to the higher energies around 1.3 eV in the
insulating phase. We have also found that the structure around 1.3 eV in
the insulating phase shows a temperature dependence similar to the
photoemission spectra and considered that this is another evidence for
the strong electron-phonon interaction.

\section{Experimental}

\begin{figure}[t]
\begin{center}
\includegraphics[width=8cm]{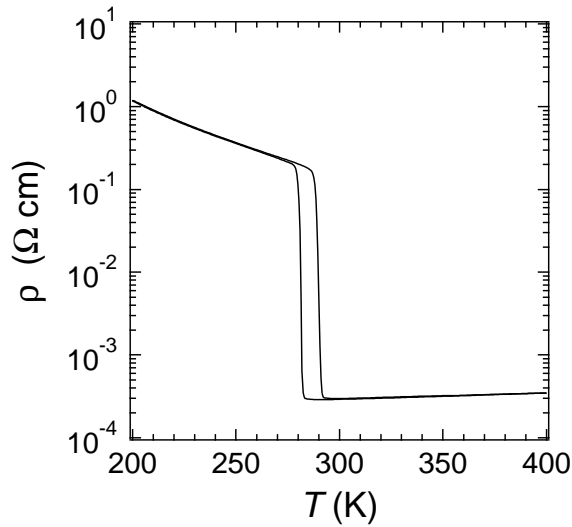} 
\caption{\label{rho} Electrical resistivity $\rho$ of VO$_2$/TiO$_2$
 (001) thin film. A jump of about three orders magnitude has been
 observed around 290 K.}
\end{center}
\end{figure}

VO$_2$ thin films were epitaxially grown on TiO$_2$(001) surfaces using
the pulsed laser deposition technique as described in
Ref.~\onlinecite{Muraoka}. The film thickness was $\sim$ 100 {\AA},
estimated by four-cycle x-ray diffraction (XRD) measurements and the MIT
was confirmed by electrical resistivity measurements (Fig.~\ref{rho}),
showing a jump of about three orders of magnitude around 290 K ($\equiv
T_{MI}$). This reduced MIT temperature of the films is due to the
compressive strain from the TiO$_2$ substrate.~\cite{Muraoka}

Near-normal incidence optical reflectivity measurements were performed
using a Fourier-type interferometer (0.006 - 1.2 eV) and a grating
spectrometer (0.8 - 6.8 eV) from 5 K to 350 K. Because VO$_2$ films have
been epitaxially grown on the TiO$_2$ (001) surface, polarization of the
incident light is perpendicular to the $c$-axis of the Rutile structure
(\mbox{\boldmath $E$}$\bot c$). As a reference mirror, we used an
evaporated Au film for the infrared regions and Ag film for the visible
region. The experimental error of the reflectivity is less than 0.5 \%
for the far- and near-IR regions and less than 1.0 \% for the mid-IR,
visible and ultraviolet regions. A detailed procedure to deduce optical
constants of VO$_2$ from the reflectivity is described bellow.

\begin{figure}[t]
\begin{center}
\includegraphics[width=8cm]{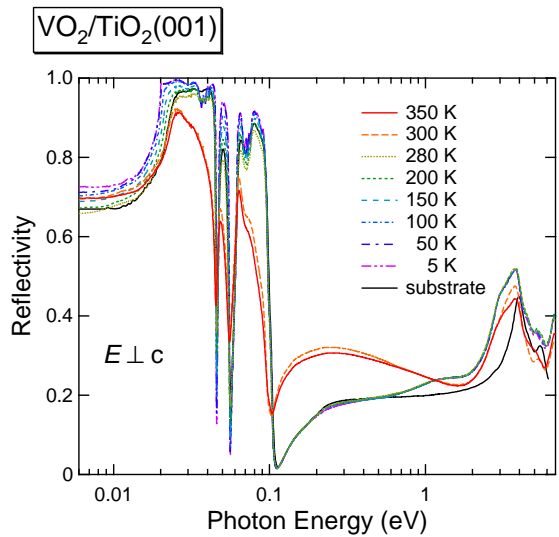} 
\caption{\label{ref} Reflectivity spectra of VO$_2$/TiO$_2$ (001) thin
 film at various temperatures. Below 0.1 eV, because the reflectivity of
 the substrate is large due to optical phonons, there is an interference
 effect between the reflection at the substrate and the film.}
\end{center}
\end{figure}

The normal incidence complex reflectivity of the two layer system can be
written as
\begin{eqnarray}
&&\hat{r}(\omega) = \sqrt{R(\omega)} \cdot e^{i\Theta(\omega)} \nonumber \\ 
&&= \frac
{\hat{r}_{0f}+\hat{r}_{fs}e^{2i\delta_f}+\hat{r}_{0f}\hat{r}_{fs}\hat{r}_{s0}e^{2i\delta_s}+\hat{r}_{s0}e^{2i(\delta_f+\delta_s)}}
{1+\hat{r}_{0f}\hat{r}_{fs}e^{2i\delta_f}+\hat{r}_{fs}\hat{r}_{s0}e^{2i\delta_s}+\hat{r}_{0f}\hat{r}_{s0}e^{2i(\delta_f+\delta_s)}},
\nonumber \\
\end{eqnarray}
where
\begin{eqnarray*}
\hat{r}_{0f}&=&[(n_f+ik_f)-1]/[(n_f+ik_f)+1], \\
\hat{r}_{fs}&=&[(n_s+ik_s)-(n_f+ik_f)]/[(n_s+ik_s)+(n_f+ik_f)], \\
\hat{r}_{s0}&=&[1-(n_s+ik_s)]/[1+(n_f+ik_f)], \\
\delta_f&=&2\pi i(n_f+ik_f)d_f/\lambda, \\
\rm{and}&&\\
\delta_s&=&2\pi i(n_s+ik_s)d_s/\lambda.
\end{eqnarray*}
$n_f, k_f, d_f, n_s, k_s,$ and $d_s$ are the refractive index, the
extinction coefficient, and the film thickness of the film and the
substrate, respectively and $\lambda$ is the wave length of the incident
light.

First, we have measured the reflectivity of the TiO$_2$ substrates at
various temperatures, and then obtained optical constants $n_s$ and
$k_s$ using Kramers-Kronig (K-K) transformation. For the extrapolation
to perform K-K transformation, we have assumed a constant value below
0.006 eV and a power-low behavior ($\propto \omega^{-\alpha}, 0 < \alpha
< 4$) above 6.8 eV. The value of $\alpha$ was adjusted so that the
obtained optical constants reproduced the reported
values~\cite{Handbook} at 300K. ~\cite{Wooten} Next, we have measured
the reflectivity $R(\omega)$ of the VO$_2$/TiO$_2$ thin films as shown
in Fig.~\ref{ref} and deduced the phase shift $\Theta(\omega)$ using K-K
transformation. We have adopted a same assumption with the TiO$_2$
substrate for the extrapolation. At last, we have obtained the optical
constants $n_f$ and $k_f$ by numerically solving the equations obtained
from the real and imaginary part of Eq. (1) multiplied by the
denominator of the right-hand side. When solving the equations, we have
neglected the terms proportional to sin$(2{\pi}n_sd_s/\lambda)$ and
cos$(2{\pi}n_sd_s/\lambda)$. Because the substrate thickness $d_s \sim$
0.5 mm is much larger than the wave length $\lambda$ and is not uniform
in the scale of $\lambda$, these terms can be replaced with their mean
value zero.~\cite{Nilson} 

\section{Results and discussion}

\begin{figure}[t]
\begin{center}
\includegraphics[width=8cm]{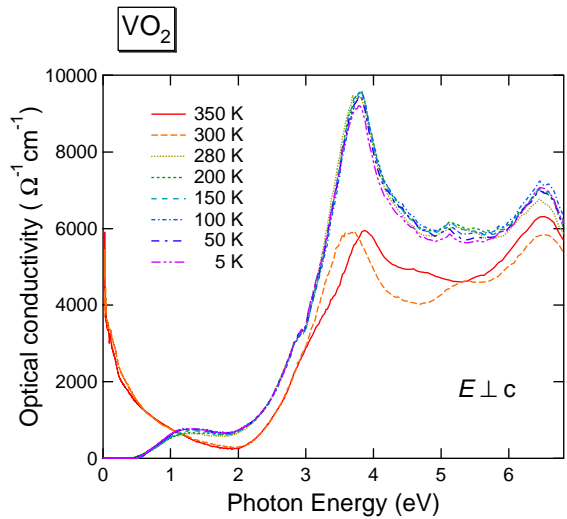} 
\caption{\label{sigma} Optical conductivity spectra of VO$_2$ at various
 temperatures deduced from the reflectivity spectra of
 VO$_2$/TiO$_2$(001) thin film.}
\end{center}
\end{figure}

Figure~\ref{sigma} shows the optical conductivity spectra of VO$_2$ at
various temperatures. The MIT is clearly observed as a spectral weight
transfer of the broad Drude-like structure above $T_{\rm MI}$ to the
Gaussian-like peak structure around 1.3 eV below $T_{\rm MI}$. The
structure around 1.3 eV can be assigned as a Mott-Hubbard excitation,
while an intense peak around 3.8 eV, which is observed both above and
below $T_{\rm MI}$, can be assigned as a charge-transfer
excitation.~\cite{Arima} In the metallic phase, the value of the
optical conductivity in the low energy region is roughly in accordance
with the DC conductivity ($\sim$ 3500 $\Omega^{-1}$cm$^{-1}$ at 300 K).

\begin{figure}[t]
\begin{center}
\includegraphics[width=8cm]{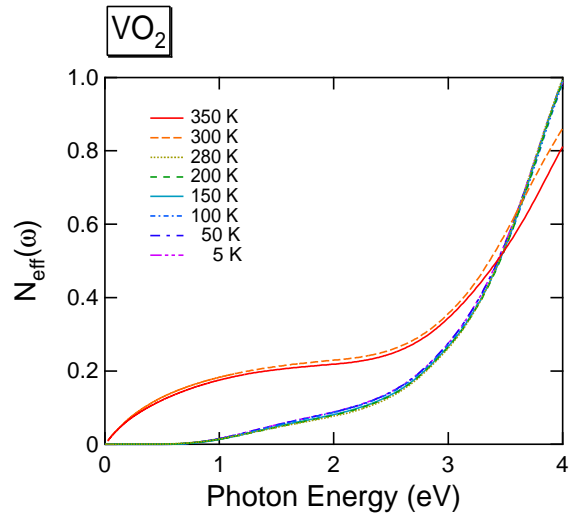} 
\caption{\label{Neff} Effective carrier number per formula unit
 $N_{\rm eff}$ of VO$_2$ obtained by integrating the optical conductivity.}
\end{center}
\end{figure}

In Fig.~\ref{Neff}, the number of the effective carriers ($N_{\rm
eff}(\omega)$) defined as
\begin{equation}
N_{eff}(\omega)\equiv\frac{2m_0 V}{\pi e^2}\int^\omega_0\sigma(\omega^\prime)d\omega^\prime
\end{equation}
is shown, where $m_0$ is the bare electron mass and $V$ is the cell
volume of the one formula unit. From $N_{\rm eff}$, the spectral weight
of the Drude-like component at 300 K can be estimated as 0.23 per
formula unit. From this value, the plasma frequency $\omega_p$ and the
effective mass $m^*$ can be estimated using a so-called $f$-sum rule,
\begin{equation}
\label{sum}
\int^{\omega_0}_0\sigma(\omega)d\omega=\frac{1}{8}\omega_p^2,
\end{equation}
where $\omega_p$ is defined as $\omega_p^2=4\pi ne^2/m^*$, $\omega_0$ is
taken as the energy for the upper limit of the contribution of the
Drude-like component (= 2.0 eV), and $n$ is the charge density (= 1/V
i.e., one electron per formula unit). Thus, $\omega_p$ and $m^*/m$ are
estimated as 3.3 eV and 4.3, respectively. These values agree well with
those estimated from the peak position of the energy-loss function
Im(-1/$\epsilon(\omega)$) by assuming its peak position corresponding to
$\omega_p/\sqrt{(\epsilon_{\infty})}$,~\cite{loss} where
$\epsilon_{\infty}$ is the optical dielectric constant and estimated as
$\sim$ 8 from Re($\epsilon(\omega)$) at $\omega$ = 2.0 eV. The spectra
of Im(-1/$\epsilon(\omega)$) at 300 K has a peak at 1.15 eV and the
$\omega_p$ is estimated as 3.25 eV. From this consistency, we can say
that the obtained mass enhancement factor $m^*/m_0$ should be reasonable
one. For another vanadium oxide $d^1$ system Sr$_{1-x}$Ca$_x$VO$_3$,
which is metallic at all temperatures, $m^*/m_0$ has been estimated as
$\sim$ 3.1 for CaVO$_3$ and $\sim$ 2.7 for SrVO$_3$ from optical
measurements.~\cite{Makino} Hence, $m^*/m_0$ of VO$_2$ is somewhat
larger than that of Sr$_{1-x}$Ca$_x$VO$_3$. Hence, we can say that the
quasiparticle renormalization or the electron correlation effect in
VO$_2$ is stronger than Sr$_{1-x}$Ca$_x$VO$_3$. This is consistent with
the quasiparticle renormalization factor $Z$ estimated from
photoemission spectroscopy. They have been estimated as $\sim$ 0.3 for
VO$_2$~(Ref.\onlinecite{PES}) and $\sim$ 0.5 for
Sr$_{1-x}$Ca$_x$VO$_3$~(Ref.\onlinecite{Sekiyama}), respectively.

\begin{figure}[t]
\begin{center}
\includegraphics[width=8cm]{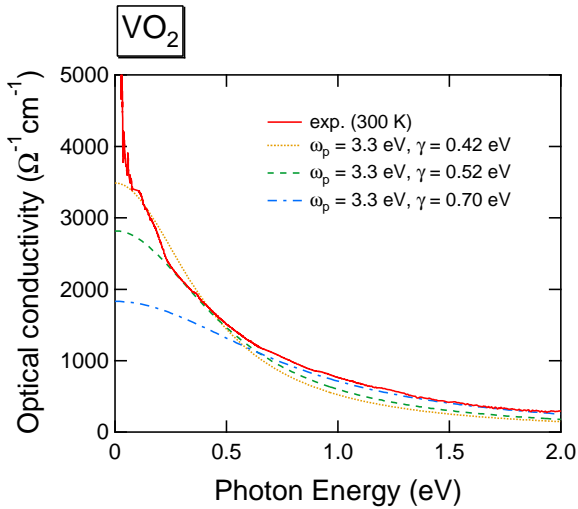} 
\caption{\label{Drude} Comparison of the optical conductivity of VO$_2$
 in the metallic phase with the simple Drude model.}
\end{center}
\end{figure}

\begin{figure}[t]
\begin{center}
\includegraphics[width=8cm]{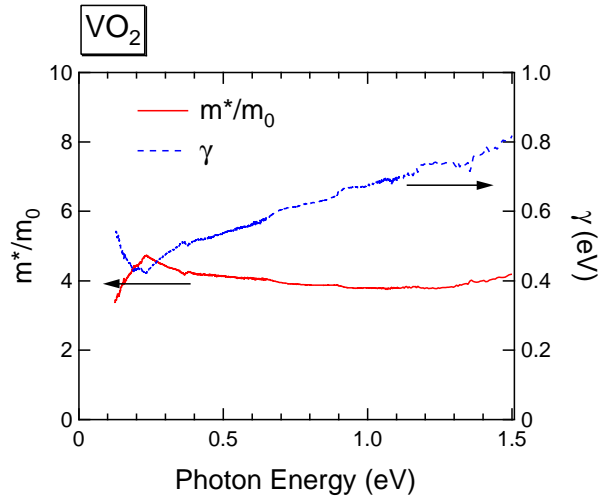} 
\caption{\label{exDrude} Energy-dependent mass enhancement factor
 $m^*(\omega)/m_0$ and the scattering rate $\gamma(\omega)$ deduced from
 the extended Drude model.}
\end{center}
\end{figure}

Using the value of $\omega_p$ estimated above, we have compared the
optical conductivity at 300 K with the simple Drude model with various
$\gamma$ (scattering rate) in Fig.~\ref{Drude}. In the simple Drude
model, the optical conductivity is given by $\sigma(\omega) = \omega_p^2
\gamma / 4\pi (\omega^2 + \gamma ^2)$. With a larger value of $\gamma$,
the simple Drude model seem to fit in the higher energy region. However,
it cannot reproduce the whole line shape of the optical conductivity
with the only one value. This should correspond to the energy dependence
of the $\gamma$.  Then, we introduce the energy dependence in $\gamma$
and $m^*/m_0$ using the extended Drude model. Figure~\ref{exDrude} shows
$\gamma(\omega)$ and $m^*(\omega)/m_0$ at 300 K above 0.1 eV. As
expected from the comparison with the simple Drude model,
$\gamma(\omega)$ increases with the photon energy and is almost
proportional to $\omega$, while the energy dependence of $m^*(\omega)/m$
is not so large. A similar behavior has been observed in other
transition-metal oxides.~\cite{Makino, Katsufuji} However, it seems to
be a still open question why $\gamma(\omega)$ is proportional to
$\omega$ rather than $\omega^2$ as expected from the standard Fermi
liquid theory.

Although the behavior of $\gamma(\omega) \propto \omega$ is commonly
observed in other transition-metal oxides, the value of $\gamma(\omega)$
of VO$_2$ is somewhat larger compared to Sr$_{1-x}$Cr$_{x}$VO$_3$. For
the case of VO$_2$, this feature may related to the temperature
dependence of the electronic resistivity.  Allen {\it et
al.}~\cite{Allen} have reported that the resistivity of the single
crystal VO$_2$ shows a linear temperature dependence and does not saturate
at least up to 840 K. They have estimated the mean free path $\sim$ 3
{\AA} at 800 K and claimed that VO$_2$ might not be a conventional Fermi
liquid. When the mean free path is such small, description for the
electronic transport by the Boltzmann equation may be not
valid. However, the unexpected small mean free path is due to the large
scattering rate and hence, these features in the electronic resistivity
and the optical conductivity may be related to the characteristic
scattering mechanism in VO$_2$

\begin{figure}[t]
\begin{center}
\includegraphics[width=8cm]{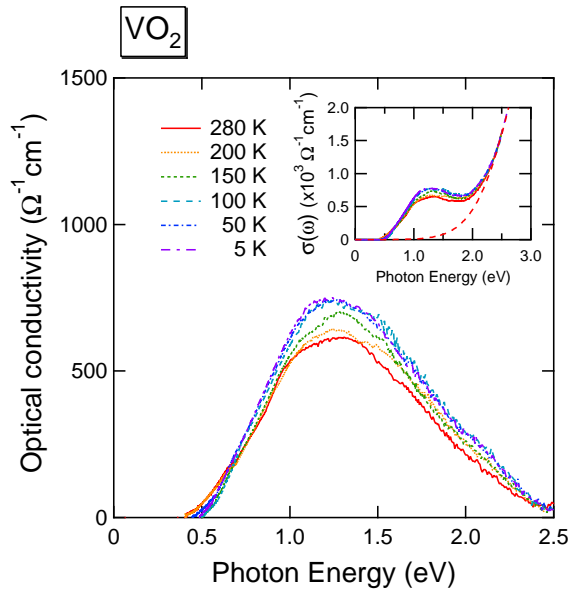} 
\caption{\label{insulating} Temperature dependence of the optical
conductivity of VO$_2$ in the insulating phase in the region of
Mott-Hubbard excitation after subtraction of the the contribution from
the higher-energy peak.}
\end{center}
\end{figure}

\begin{figure}[t]
\begin{center}
\includegraphics[width=8cm]{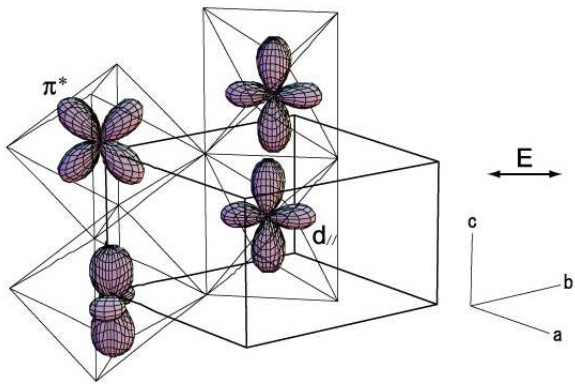} 
\caption{\label{Fig8} Configuration between the $d_{\|}$ and $\pi^*$
orbitals. The occupied $d_{\|}$ orbitals can transfer to the unoccupied
$\pi^*$ orbitals of the second-nearest-neighbor V atom with the
 \mbox{\boldmath $E$}$\bot c$ polarization, while it cannot transfer to
 the $d_{\|}$ orbitals of the nearest and the second-nearest-neighbor V
 atoms. Thick lines indicate the unit cell of the rutile structure and
 thin lines indicate oxygen octahedra. (Note that the unit cell volume
 in the insulating phase is twice of the rutile structure in the
 metallic phase.)} 
\end{center}
\end{figure}

Next, we would like to discuss the optical conductivity spectra in the
insulating phase. Figure~\ref{insulating} shows the optical conductivity
spectra of VO$_2$ in the region of the Mott-Hubbard excitation. The
contribution from the higher energy peak is subtracted by assuming as
shown in the inset. We would like to note that the peak position of this
excitation is far less than twice of the peak position of V $3d$ peak of
the photoemission spectra ($\sim$ 1.0 eV). For example, the dynamical
mean field theory (DMFT) predicts that the spectral function of the
half-filled Hubbard model has peaks at $\omega = -U/2$ and $U/2$
corresponding the lower- and upper-Hubbard band, respectively and that
the optical conductivity has a peak at $\omega = U$ in the insulating
region. In fact, Sr$_{1-x}$Ca$_x$VO$_3$ and V$_2$O$_3$ have been
interpreted within this scheme.~\cite{Makino,V2O3} For the case of
VO$_2$, the different $d$-electron bands, $d_{\|}$ and $\pi^*$
bands~\cite{Goodenough} could be related to this transition. Because the
polarization of the incident light is \mbox{\boldmath $E$}$\bot c$, the
occupied $d_{\|}$ orbitals cannot transfer to the nearest neighbor V
atoms along $c$-axis (See Fig.~\ref{Fig8}). Hence, this structure is
assigned to the $d$-$d$ transfer to the second nearest neighbor V
atoms. Furthermore, $d_{\|}$ orbitals between the
second-nearest-neighbor V atoms are orthogonal, this structure should be
the $d_{\|}$-$\pi^*$ transition between the second-nearest-neighbor V
atoms.

The spectra in Fig.~\ref{insulating} have the similar line shapes and
show a similar temperature dependence to the photoemission
spectra. Because the optical conductivity spectra in Fig.~\ref{sigma}
show almost no temperature dependence around the region from 2.5 eV to
3.0 eV in the insulating phase, the same Gaussian function was used to
subtract the contribution from the higher energy peak for all the
spectra in Fig.~\ref{insulating}. Hence, this subtraction procedure does
not affect the temperature dependence below 2.5 eV. In the recent
photoemission study,~\cite{PES} the line shape of the V $3d$ band and
its temperature dependence has been reproduced by the independent boson
model and attributed to the strong electron-phonon coupling. Using the
independent boson model, the optical conductivity at finite temperature
can be written by a similar expression to the spectral
function,~\cite{Mahan}
\begin{eqnarray*} 
\sigma(\omega) \propto \frac{\pi}{\omega} e^{-g_{if}(2N+1)}\sum_l\frac{g_{if}^l}{l!}\sum_{m=0}^l{}_l C_mN^m(N+1)^{l-m} \\
\times \delta(\omega-\varepsilon_i+\varepsilon_f+\Delta_i-\Delta_f-(l-2m)\omega_0).
\end{eqnarray*}
This describes a transition from the initial state $i$ to the final
state $f$. $N$ is the phonon occupation number, $g_{if}$ is the
effective electron-phonon coupling constant, which is dependent on the
electron-phonon coupling of both the initial and final states,
$\omega_0$ is the phonon energy, $\varepsilon_i$ and $\varepsilon_f$ are
the energies of the initial and final states, respectively, and
$\Delta_i$ and $\Delta_f$ are the electron self-energies of the initial
and final states, respectively. From this expression, similar
temperature dependence to the photoemission spectra is expected for the
optical conductivity. Hence, we conclude that the temperature dependence
of the optical conductivity is another evidence for the strong
electron-phonon interaction.

\section{Conclusion} 
We have studied the charge dynamics of VO$_2$ by the optical
reflectivity measurements. The optical conductivity clearly shows a
metal-insulator transition. In the metallic phase, the spectral weight
of the Drude-like component and the mass enhancement factor $m^*/m_0$
have been estimated as $\sim$ 0.23 and 4.3, respectively. This $m^*/m_0$
is somewhat larger than another $d^1$ vanadium oxide
Sr$_{1-x}$Ca$_x$VO$_3$. From this, we have concluded that the electron
correlation effect in VO$_2$ is stronger than
Sr$_{1-x}$Ca$_x$VO$_3$. From the extended Drude model, $\gamma(\omega)$
is rather large and almost proportional to $\omega$. This would be
related to the characteristic scattering mechanism in the metallic phase
of VO$_2$. In the insulating phase, the broad Gaussian-like structure
has been observed around 1.3 eV. This would be a $d_{\|}$-$\pi^*$
transition between the second-nearest-neighbor V atoms. This structure
show a similar temperature dependence to the photoemission spectra. We
have concluded that this is another evidence for the strong
electron-phonon coupling.

\section*{Acknowledgements}
The authors would like to thank J. Matsuno and A. Fujimori for
enlightening discussions. This work was partially supported by the 21st
Century COE program of Nagoya University and Grant-in-Aid for Scientific
Research from the Ministry of Education, Culture, Sports, Science and
Technology, Japan.


\begin{thebibliography}{99}
\bibitem{loss}
Strictly, even in the simple Drude model, only when $\gamma$ is small 
enough, the peak position of the energy-loss function is located 
at $\omega_p/\sqrt{(\epsilon_\infty)}$.
\end{thebibliography}
\end{document}